\documentstyle[aps,twocolumn]{revtex}

\begin{document}
\date{\today}

\title{Distribution of nearest distances between nodal points for the
Berry function in two dimensions}
\author{Alexander I. Saichev$^{1,2}$, Karl-Fredrik Berggren$^2$, Almas F. Sadreev$^{2,3}$}
\address{1) Department of Radiophysics, Nizhny Novgorod University, Gagarin pr.,
23, 603600, Nizhny Novgorod, Russia\\
2) Department of Physics and Measurement Technology,
Link\"{o}ping University, S-581 83 Link\"{o}ping, Sweden\\
3) Kirensky Institute of Physics, 660036, Krasnoyarsk, Russia}
\maketitle
\begin{abstract}
According to Berry a wave-chaotic state may be viewed as a superposition of
monochromatic plane waves with random phases and amplitudes.
Here we consider the distribution of nodal points associated with this state.
Using the property that both the real and imaginary parts of the wave function
 are random Gaussian fields we analyze the
correlation function and densities of the nodal points.
Using two approaches (the Poisson and Bernoulli) we derive the distribution of
nearest neighbor separations. Furthermore the distribution functions
for nodal points with specific chirality are found. Comparison is made with results from
from numerical calculations for the Berry wave function. 
\end{abstract}

\section{Introduction}
The nature of the quantum eigenstates in billiards,
which are classically chaotic, has been
subject to much theoretical and experimental work.
The seminal studies  by McDonald and Kaufmann \cite{McDonald} of the morphology
of the two-dimensional (2D) eigenstates in a closed Bunimovich stadium have
revealed characteristic complex patterns of disordered, undirectional
and non-crossing nodal lines.
The spatial behavior of the eigenstates of chaotic billiards is still of
considerable theoretical
and experimental interest. For recent theory see, e.g.,
\cite{Backer,Hu,Kole,Bies}, the review by Robnik \cite{Robnik},
and references cited; examples of measurements on electron billiards and
wave-dynamical analogues are found in, e.g., \cite{Com} (quantum corrals),
\cite{Sridhar,Alt,Alt1,Veble} (microwave cavities),
 \cite{Schaadt} (acoustic resonators),
\cite{Kudrolli} (surface water waves in tanks) and in \cite{Stock} in general.
Other well known signatures of quantum
chaos in closed billiards are related to the distribution of nearest level separations
and spectral rigidity.

For open billiards, i.e., billiards with attached leads, the picture is less clear. One may use the
poles of the scattering matrix which are related to the decay
time from a billiard \cite{Fyodorov,Dittes,Ishio}.   When
 transport through a billiard takes place one may as an alternative
  focus on the fact that the wave function $\psi$
  is  scattering state with both real and imaginary parts. If we restrict ourselves
 to 2D systems, as we will do throughout this work, this means that there are
 two separate sets of nodal lines at which either $Re[\psi]$ or $Im[\psi]$
  vanish. The intersections of the two sets at which $Re[\psi] = Im[\psi] =0$  define the
  nodal points. Numerical simulations have shown that the shape of
distribution function for the nearest distances
  between these nodal points (DFNDNP) depends on whether the
  billiard  is nominally either regular or
  irregular \cite{Berg1}. For transmission through
  chaotic billiards the DFNDNP appears to have a general characteristic form, while for regular billiards
  like, for example, rectangular ones there are specific features of the DFNDNP
  that depend on the particular geometry, at least as long as only a few channels are open.
  Thus, besides the vivid physical role played by the nodal points as centers
  of vortical motion
\cite{hirsch,hirsch1,Berg3,wu,exner,stoeck}, their statistical distribution may
 tell if chaos is present or not. The present work relates to quantum 
 transport transport in open electron billiards. The issue of wave function 
 singularities is, however, part a much broader context 
 \cite{Berry3,Berry4,conf,soskin}.

  An appealing argument that favors our view that DFNDNP serves as a
   signature of quantum chaos
is the coincidence with the corresponding
distribution function for the Berry state \cite{Berg1}. According to Berry's conjecture
a chaotic state can be viewed as a
 superposition of a large number of interfering monochromatic de Broglie
waves\cite{Berry3}
\begin{equation}\label{berry}
    \psi({\bf r})=\sum_j a_j\exp(i{\bf k}_j\cdot{\bf r}+\phi_j)
\end{equation}
    where $a_j$ and $\phi_j$ are independent random variables and ${\bf k}_j$
    are randomly oriented wave vectors of equal length.
The Berry wave function may be regarded as a standard measure of quantum chaos.
In fact, there are
  beautiful experimental observations of Berry waves on the surface of water in an agitated ripple tank
  with stadium-shaped walls \cite{blumel}.

\begin{figure}
\includegraphics{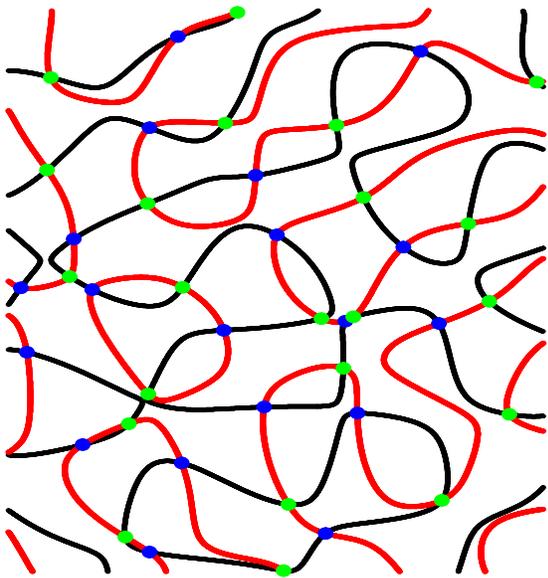}
\caption{Typical pattern of nodal lines $Im[\Psi(x,y)]=0$ (black lines) 
and $Re[\Psi(x,y)]=0$ (red lines)
for the Berry function. Nodal lines in each set do not cross. 
Points at which the two sets intersect are the nodal points around which there 
is vortical flow in either clockwise (green dots) or anti-clockwise 
(blue dots) direction. }
\label{fig1}
\end{figure}

 So far all our conclusions about DFNDNP rest on numerical experiments \cite{Berg1}.
 The Berry state is, however,
available in a mathematical form that invites to analytic approaches.
In the present work we will therefore model
the DFNDNP $f(r)$ and its main
asymptotic behavior analytically using the fundamental property that the Berry
 function (1) is Gaussian
random field \cite{Berry3}.
We will also show that there are other types of distribution functions that are
related to the
chirality of the nodal points. Each nodal point is a topological singularity of the
wave function\cite{hirsch,hirsch1,wu,exner,stoeck,Berry3}. As a result
 there is  a vortex centered around each nodal point with definite
chirality depending on whether the current flows
clockwise or anti-clock wise as indicated in Fig, \ref{fig1}.
\vspace*{9cm} 

We therefore label each nodal point by
$\sigma =\pm 1$. In analogy with  $f(r)$ we therefore
introduce the distribution functions $f_{\sigma,\sigma'}(r)$ for nearest
 neighbor separations between points with chiralities $\sigma$ and $\sigma'$.
Analytic expressions for these distributions will be derived below and
compared with numerical computations using the Berry wave function.
As will be pointed out in the text our results partly overlap with recent 
work by Berry and
Dennis \cite{Berry3} (the pair correlation functions $g_{\sigma,\sigma'}$ 
and the relation between the mean
density and wave number $k$). Most recently Dennis has also considered 
the distribution of nearest distances among nodal points using the 
Poisson model\cite{dennis}. 

In the following sections we will derive expressions for the distributions 
of nearest neighbor separations between points with chiralities $\sigma$ 
and $\sigma'$. For this purpose we will also have to consider the pair 
correlation functions $g_{\sigma,\sigma'}$. We will make use of two different 
analytic approaches based on the Poisson and Bernoulli models. In addition we
 will also
calculate the distributions by direct numerical methods, i.e., we locate the 
nodal points by simply computing the nodal lines  for $Im(\Psi)$ and 
$Re(\Psi)$ and how they cross. In principle the numerical results represent the correct distributions and gives us a way to test the accuracy of the different analytic approaches.

\section{Definition of variables}

Consider the Berry function (\ref{berry}) as the complex random function
\begin{equation}\label{1.1}
  \psi({\bf r})=u({\bf r})+i v({\bf r})
\end{equation}
where ${\bf r}$ is the 2D position with Cartesian coordinates
$x_1,x_2$ and $u({\bf r}),$ $v({\bf r})$ are two real random
fields. We  assume that $u({\bf r})$ and $v({\bf r})$ are
mutually statistically independent, homogeneous and
isotropic Gaussian random fields with zero mean. The 
correlation function has the well known form
\begin{equation}\label{bessel}
a(s)=\langle u({\bf r}) u({\bf r+ s})\rangle=
\langle v({\bf r}) v({\bf r+ s})\rangle = J_0(ks).
\end{equation}
which is a direct consequence of the Berry function (\ref{berry}).
To find the statistical properties of the nodal points ${\bf r}_j$
we have to consider the intersections of the zero level curves (nodal lines)
of the fields $u({\bf r})$ and $v({\bf r}),$ i.e., the roots of the two
equations:
$$
u({\bf r}_j)=v({\bf r}_j)=0 \quad ({\bf r}_j\in {\bf R}^2)\,.
$$

As mentioned in the introduction the  nodal points are the centers of vortices.
The associated probability current ${\bf J( r)}$  is proportional to
\begin{equation}\label{1.2}
  {\bf J(r)}=\text{Re}\,
  \left[\psi^*({\bf r})\,i\nabla\psi({\bf r})\right]=
  v({\bf r})\nabla u({\bf r})-u({\bf r})\nabla v({\bf r})\,.
\end{equation}
More rigorously the nodal points are responsible for the vortices in
the sense that the loop integral \cite{hirsch,wu,exner,stoeck}
$$
\oint d{\bf r v} =\oint \nabla\theta d{\bf r}
$$
must be non-zero only if it encloses a nodal point. Here $\theta$ is the phase
of the complex wave function. In the present work we consider the vorticity field
\begin{equation}\label{omega}
\mbox{\boldmath $\omega $}=\nabla\times {\bf J}\,.
\end{equation}
In our two-dimensional case it is normal to the $(x,y)$-plane, i.e.,
$\mbox{\boldmath $\omega $}({\bf r})=\omega ({\bf r})\, 
\mbox{\boldmath $\hat{n}$}_z$
where $\mbox{\boldmath $\hat{n}$}_z$ is 
the normal unit  vector. 
Substituting
(\ref{1.2}) into (\ref{omega}) we have
\begin{equation}\label{1.5}
\mbox{\boldmath $\omega$}({\bf r})=
\left[\nabla u({\bf r})\times\nabla v({\bf r})\right].
\end{equation}
At the nodal point ${\bf r}_j$
$$
\omega_j=\omega({\bf r}_j)
$$
is the angular velocity of the current in the very vicinity of ${\bf r}_j$.
 In the following we will call  $\omega_j$  the
{\it vorticity} of the $j$-th nodal point.
\section{General formulas for the density of nodal points}
If we introduce the density of nodal points as
\begin{equation}\label{density}
{\cal R}({\bf r})=|\mbox{\boldmath $\omega$}({\bf r}_j)|\delta(u({\bf r}))\delta(v({\bf r}))
\end{equation}
we obtain
\begin{equation}\label{3.0}
{\cal R}({\bf r})=\sum_j \delta({\bf r}- {\bf r}_j).
\end{equation}
Let us introduce also another singular function
\begin{equation}\label{3.1}
  {\cal R}_v({\bf r})=
  \omega({\bf r})\delta(u({\bf r}))\delta(v({\bf r}))=
  \sum_j \sigma_j \delta({\bf r- r}_j)
\end{equation}
where
\begin{equation}\label{3.2}
\sigma_j=\frac{\omega_j}
{|\omega_j|}
\end{equation}
equals  $\pm1$ for clockwise and anti-clockwise vorticities $\omega_j$, respectively.
 Therefore  (\ref{3.2}) defines the
sign of the vorticity of the nodal points. Below we will refer to $\sigma$ as 
chirality. In ref. \cite{Berry3} it is named topological charge. There are as many 
points with $\sigma=1$ as with $\sigma=-1$. 

Next, let us define the mean density
\begin{equation}\label{4.2}
\langle{\cal R}({\bf r})\rangle=\rho
\end{equation}
and the correlation function for the random density
\begin{equation}\label{4.3}
  G(s)=\langle{\cal R}({\bf r}){\cal R}({\bf r- s})\rangle=
    \left<\sum_{i,j}
  \delta({\bf r- r}_i)\delta({\bf r- r}_j-{\bf s})\right>.
\end{equation}

Notice that because ${\cal R}({\bf r})$ is a statistically
homogeneous and isotropic random field, the mean density $\rho$ is
constant and the correlation function $G(s)$ depends only on the distance
$s$ between the points of observation ${\bf r}$ and ${\bf r- s}$.
Formulas (\ref{density}), (\ref{3.1}), (\ref{4.2}) and (\ref{4.3}) 
form the basis of the statistical analysis
of the nodal points distribution assuming that the functions $u({\bf r})$ and
$v({\bf r})$ are  random functions.

The mean density defines a characteristic scale
\begin{equation}\label{scale}
  s_\rho=\frac{1}{\sqrt{\rho}}
\end{equation}
which we will use below as the natural unit of distances in
the ``gas'' of randomly distributed points, i.e., we will use
the dimensionless variable
\begin{equation}\label{4.4}
\ell=\frac{s}{s_\rho}=\sqrt{\rho}\,s\,.
\end{equation}
Moreover it is convenient to formulate some analytical results in terms of
the dimensionless pair pair correlation function with dimensionless argument
\begin{equation}\label{g}
  g(\ell)=\frac{1}{\rho^2}\,
  G\left(\frac{\ell}{\sqrt{\rho}}\right)\,.
\end{equation}

We now introduce the mean density $\gamma(r)$ around of a given point.
One can show that
\begin{equation}\label{4.5}
  \gamma(s)=\frac{1}{\rho}\;G(s).
\end{equation}
which will play a crucial role
in the following derivation of DFNDNP for the Berry function.

We now consider some useful relations for the statistics of the nodal points.
Firstly, consider the mean number of points inside a circle $C_r$ with
radius $r$ centered at some given point. It is obvious that the mean
number of points enclosed by the circle is equal to
\begin{equation}\label{4.6}
  \langle n(r)\rangle=2\pi\int_0^r \gamma(y)y\,dy.
\end{equation}
Using the dimensionless coordinate (\ref{4.4}) one obtains
\begin{equation}\label{4.7}
  \langle n(\ell)\rangle=2\pi\int_0^\ell g(r)r\,dr .
\end{equation}
This relation takes into account that the dimensionless correlation function
$g(r)$ is at the same time the dimensionless mean density.
Secondly, consider also the relation for the mean number of nodal points
$$
 \langle n(\ell)\rangle=\sum_{n=1}^\infty n\text{P}(n;\ell).
$$
Here $\text{P}(n;\ell)$ is the probability that $n$
neighboring points belong to the circle.
These probabilities satisfy the normalization condition
\begin{equation}\label{4.8}
  \text{P}(0;\ell)=1-\sum_{n=1}^\infty\text{P}(n;\ell)\,.
\end{equation}

The probability $P(0;\ell)$ is of great importance because it is
directly related to the distribution function for nearest
distances between a given point and its neighbouring points
(DFNDNP) $f_{\min}(\ell)$.
This is because the cumulative distribution of the nearest distances
$\ell_{\min}$ is given by
\begin{equation}\label{4.9}
  F_{\min}(\ell)=\text{P}(\ell_{\min}<\ell)=
  1-\text{P}(0;\ell).
\end{equation}
Therefore we may now write the following relation for the dimensionless
distribution of nearest distances
\begin{equation}\label{4.10}
f_{\min}(\ell)=-\frac{\partial}{\partial\ell}\text{P}(0;\ell)\,.
\end{equation}
Thus, the last formula reduces the problem of calculating the DFNDNP
to that of finding $\text{P}(0;\ell)$. Below we will find
approximate expressions for $\text{P}(0;\ell)$.
However, we will first present
asymptotic formulas for $\text{P}(0;\ell)$ and
the DFNDNP. For small
$\ell$ one may replace the exact
relations (\ref{4.7}), (\ref{4.8}) with the asymptotic forms
$$
  \langle n(\ell)\rangle\sim \text{P}(1;\ell)\,,\quad
  \text{P}(0;\ell)\sim 1-\text{P}(1;\ell)\quad \ell\to 0.
$$
Finally, from (\ref{4.7}), (\ref{4.10}) and the last relation above one obtains
the folowing asymptotic formula for the exact DFNDNP
\begin{equation}\label{25}
  f_{\min}(\ell)\sim\frac{\partial}{\partial \ell}
  \langle n(\ell)\rangle=2\pi\ell\,g(\ell)\quad
  (\ell\to 0)\,.
\end{equation}

Let us now apply these general considerations to the Berry function (\ref{berry}).
First of all we will calculate the mean density
$\rho$ (\ref{4.2}).
Using the definition (\ref{density}) and the fact that the variables of 
the homogeneous
Gaussian field and its derivatives are statistically independent
at the same point we have
\begin{equation}\label{rho1}
\rho=\langle |\omega({\bf r})|\rangle\,
\langle \delta(u({\bf r}))\rangle
\langle\delta(u({\bf r}))\rangle\,.
\end{equation}
It is straightforward to show that
\begin{equation}\label{deltau}
\langle\delta (u({\bf r}))\rangle\langle\delta (v({\bf r}))\rangle=
\frac{1}{2\pi}, \quad  \langle|\omega|\rangle
=\frac{k^2}{2}
\end{equation}
where $k$ is modulus of wave vector of the Berry function (\ref{berry}).
Therefore, substituting (\ref{deltau}) into (\ref{rho1}) we obtain the final
expression for the mean density
\begin{equation}\label{rho2}
\rho=\frac{k^2}{4\pi}.
\end{equation}
This exact relation has been derived recently also by Berry and Dennis \cite{Berry3}

\begin{figure}
\includegraphics{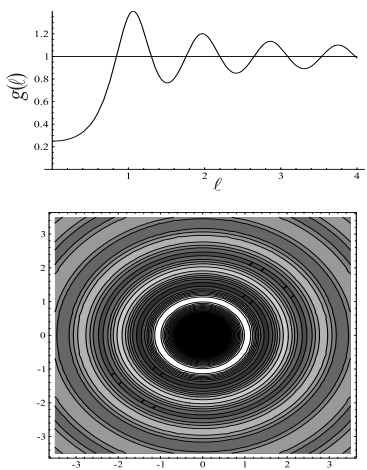}
\caption{The dimensionless correlation function $g$  nodal for the Berry
wave function (\ref{berry}) versus the dimensionless radius $\ell$.}
\label{fig2}
\end{figure}

Secondly, consider the density correlation function
(\ref{4.3})
\begin{equation}\label{dcf}
G(s)=\langle|\omega({\bf r})\omega({\bf r+s})|
\delta(u({\bf r}))\delta(u({\bf r+ s}))
\delta(v({\bf r}))\delta(v({\bf r+ s}))
\rangle\,.
\end{equation}
The calculation of $G(s)$ is given in the Appendix (\ref{61}).
and the dimensionless pair correlation  function for arbitrary $\ell$ is
plotted in Fig. \ref{fig2}.
\vspace*{7cm} 

The general behavior reminds superficially of the 
correlation functions for 
amorphous solids with short range order and distinct shell-like structures present. 
The correlations are, however, more long-range in the present case. 
The same pair correlation function was recently
obtained by Berry and Dennis \cite{Berry3}, although expressed in a different
analytic form. The derivations are somewhat tedious as indicated by the Appendix.
It is therefore rewarding that there is perfect numerical agreement with Berry 
and Dennis' results\cite{Berry3}.

Using the expressions in the appendix we can find the asymptotic expression
\begin{equation}\label{asymdcf}
  g(\ell)\sim 1/4 \quad (\ell\to 0)
\end{equation}
from which we obtain the asymptote for the DFNDNP (\ref{25})
at small $\ell$
\begin{equation}\label{asymdfnd}
  f_{\min}(\ell)\sim \frac{\pi}{2}\;\ell
  \quad (\ell\to 0)\,.
\end{equation}
This exact result is useful for testing approximate analytical
solutions and numerical simulation data.


\section{The Poisson and Bernoulli approximations for the DFNDNP}
In order to model the DFNDNP by analytic means we may in
a first attempt use the Poisson approximation. This approach has been discussed recently also by Dennis\cite{dennis}.
The Poisson approximation implies that all points around a
given one (which is located in center) are statistically independent, i.e., it neglects higher order correlations.
Therefore we have to take into account  correlations only between
the given point and its neighbours. These correlations can
be incorporated using the mean density of points $\gamma$ around the
given point
(\ref{4.5}). According to the Poisson law the probability that no
other points belong to circle
with dimensionless radius $\ell$ is
\begin{equation}\label{P0}
P(0,\ell)=\exp(-\langle n(\ell) \rangle)=
\exp\left(-2\pi\int_0^\ell z\,g(z)\,dz\right)\,.
\end{equation}
Using the relation (\ref{4.10}), we easily obtain the formula for the
DFNDNP in the Poissonian approximation
\begin{equation}\label{dfnd2}
  f_{\min}(\ell)\approx 2\pi\ell\,g(\ell)\,
  \exp\left(-2\pi\int_0^\ell z\,g(z)\,dz\right)\,.
\end{equation}
One notes that for
small $\ell$  the asymptote of the approximate DFNDNP (\ref{dfnd2}) coincides
with exact one (\ref{asymdfnd}).

For the special case of uniformly distributed and completely random
points $(g(\ell) = 1)$ we immediately obtain the well known result \cite{haake,Eggert}
\begin{equation}\label{eggert}
  f_{\min}(\ell)=2\pi\ell\exp(-\pi\ell^2)\,.
\end{equation}

\begin{figure}
\includegraphics{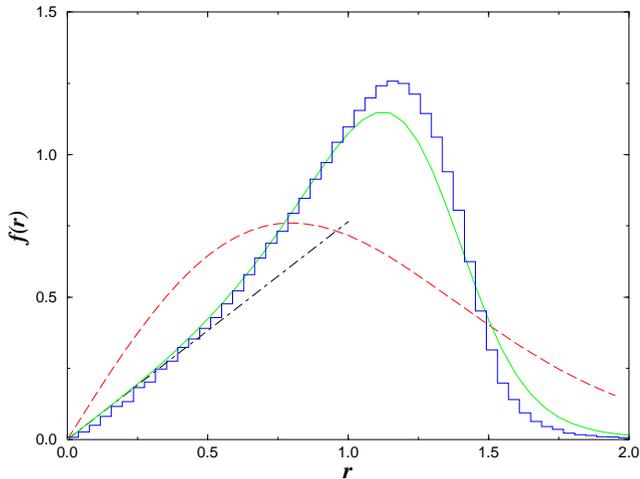}
\caption{DFNDNP versus dimensionless
distance  $r=\ell/\langle\ell\rangle$ for (a) random points (\ref{eggert})
(dashed curve) and  (b) for the Poissonian approximation (\ref{dfnd2}) (solid 
curve) with $<\ell>=0.657$. The straight dash-dot line is the corresponding
asymptote (\ref{asymdfndz}). The histogram shows the distribution
as obtained from direct numerical calculations of the positions of
the nodal points $Re(\Psi(x_j,y_j)=Im(\Psi(x_j,y_j)=0$ for the Berry function
(\ref{berry}). In these simulations we have generated the nodal points in a
large number of samples, typically of size
$(60\times 60)$ and with $k=\sqrt{2\pi}$. The number of random plane waves 
included has ranged from $20$ to $80$. In the example shown we have included 
40 plane waves and averaged over 200 samples. Except for statistical variations
the same results are obtained also for other choices of the number of plane 
waves, sample size and value of $k$.}
\label{fig3}
\end{figure}

For convenience we also introduce the
new dimensionless radius
\begin{equation}\label{32}
  r=\ell/\langle\ell\rangle
\end{equation}
which refers to mean distance between nearest nodal points $\langle\ell\rangle$.
The main asymptotic for the DFNDNP (\ref{asymdfnd}) then reads
\begin{equation}\label{asymdfndz}
  f(r)\sim \langle\ell\rangle^2\,\frac{\pi}{2}\, r=\nu r.
\end{equation}
  A comparison between (\ref{dfnd2}),
(\ref{eggert}) and the numerically calculated DFNDNP are given in 
Fig. \ref{fig3}.
\vspace*{7cm} 
 
Obviously the simplest model with $(g(\ell) = 1)$ cannot reproduce the true DFNDNP
for the simple reason that the nodal points are not  random points. The Poisson
approximated function (\ref{dfnd2}) is obviously in much better agreement with the numerical results
although the distribution falls off to quickly at large separations.
The agreement for small
$z$ is more satisfactory with $\nu=0.765$ which is rather close to
the value $0.68$
obtained from the direct numerical calculations. Although the Poissonian modeling
gives reasonable results we need to go beyond it for a better description of the
intrinsic statistical, higher-order correlations among the nodal points as 
indicated by Fig. \ref{fig2}.

A general disadvantage of the Poissonian approach is that all nodal points are
competing with each other to be neighbours of a given point. It is clear,
however, that
only nearest neighbours of the given point actually
participate in such a competition. Therefore we
consider the Bernoulli approximation for the nearest distances of
points which takes into account the competition between 
neighbouring points.  Similar to the Poisson approximation we again consider
the circle $\cal C_R$ of radius $R$ with the center at
a given point $\cal O$ and assume that all points except the given one
are statistically independent.
Furthermore, let us assume that the total number of points inside
the circle $\cal C_R$  is just equal to the mean density integral
\begin{equation}\label{35}
  n(R)=2\pi\int_0^R g(\ell)\ell\,d\ell\,.
\end{equation}
With these conditions the distribution of each randomly located
point belonging to $\cal C_R$ point  is exactly equal to
\begin{equation}\label{36}
  f(\ell)=\frac{g(\ell)}{n(R)}\,.
\end{equation}
Next, let us  consider another circle ${\cal C}_\ell$ with
the center at the same origin ${\cal O} $ and radius $\ell<R$. Obviously the
probability to find a point in this smaller circle
is equal to
\begin{equation}\label{37}
  p(\ell)=\int\limits_{\cal C_\ell} f({\bf\ell})\,d^2\ell=
  \frac{\langle n(\ell)\rangle}{n(R)}
\end{equation}
where
\begin{equation}\label{38}
  \langle n(\ell)\rangle=2\pi\int_0^\ell g(r)r\,dr\,.
\end{equation}
In the same way we have that the probability that a point does not fall into the
circle $\cal C_\ell$ is equal to
\begin{equation}\label{39}
  q(\ell)=1-p(\ell)=1-\frac{\langle n(\ell)\rangle}{n(R)}\,.
\end{equation}
Since points are  statistically independent the probability for
all points to be outside the circle $C_\ell$  equals
\begin{equation}\label{40}
  \text{P}(0;\ell)=
  \left(1-\frac{\langle n(\ell)\rangle}{n(R)}
  \right)^{n(R)}\,.
\end{equation}
From eqns. (\ref{4.9}), (\ref{4.10}), and (\ref{35}) it  follows directly
that
\begin{equation}\label{41}
  f_{\min}(\ell)=\frac{\partial}{\partial\ell}\text{P}(0;\ell)=
  2\pi \ell \, g(\ell)
  \left(1-\frac{\langle n(\ell)\rangle}
  {n(R)}\right)^{n(R)-1}.
\end{equation}
To obtain  the DFNDNP analytically we make the following
approaches within the Bernoulli approximation. (i) In formula (\ref{41}) we replace
the number of points $n(R)$ by the mean number
of points
  $\langle n(R)\rangle$ (\ref{38}).
(ii) We choose the radius $R$ in such a way that inside the circle $C_R$ there
are a certain number of points which compete with each other to be the nearest
neighbour to
 $\cal O$. The minimum number of points in the circle is obviously
three if we also include the central point. In the following we will simply use 
this value.
As a result we obtain, instead of formula (\ref{60}), the following expression
\begin{equation}\label{42}
  f_{\min}(\ell)\cong 2\pi \ell g(\ell)
  \left(1-\frac{\langle n(\ell)\rangle}
  {3}\right)^2\,.
\end{equation}
It is easy to verify that this approximate distribution is normalized
and has the same asymptote as the exact distribution function
(\ref{asymdfnd}).
\vspace*{7cm} 
\begin{figure}
\includegraphics{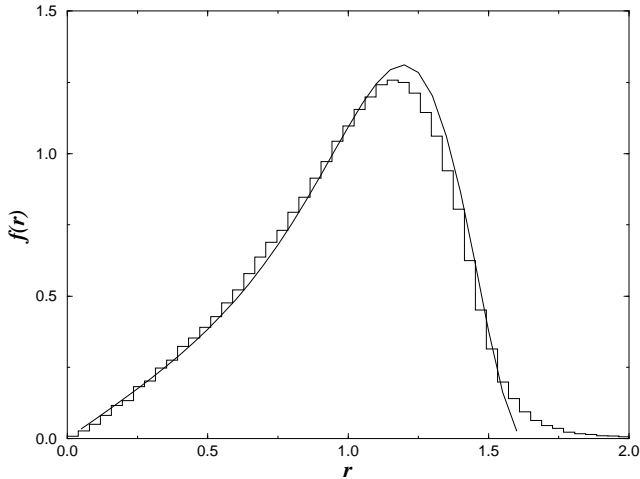}
\caption{Plots of the DFNDNP for the Berry function (\ref{berry}) versus
dimensionless distance (\ref{32}) with $<\ell>=0.658$.
Solid curve is the Bernoulli approximated distribution (\ref{42}). The histogram
is the same as in previous figure.}
\label{fig4}
\end{figure}

In Fig. \ref{fig4}
the analytic results in (\ref{42}) are compared with  the numerical
 distribution obtained directly from the Berry function.
The Bernoulli approach is evidently more powerful
than our previous attempt in predicting the DFNDNP (\ref{dfnd2}). The reason
would be that we now catch some of the higher-order correlations. 

The distribution
(\ref{42}) has the same linear behaviour at small $\ell$ as in previous
expressions (\ref{asymdfnd}) and (\ref{asymdfndz}). The coefficient
$\nu=0.678$ which is quite close to numerics ($0.68$).
\section{The mean chiral densities and correlation functions}

To gain more detailed statistics of nodal points we consider
statistical characteristics of chiral dependent
nodal points density similar to (\ref{3.1})
\begin{equation}\label{Rsigma}
{\cal R}_{\sigma}({\bf r})=\sum_{j_{\sigma}} \delta({\bf r}- {\bf r}_{j_{\sigma}}).
\end{equation}
where $j_{\sigma}$ numerates positions of vortices with chirality $\sigma=\pm 1$.
Formulas (\ref{3.0}) and (\ref{3.1}) can be written via the chiral densities
(\ref{Rsigma}) 
\begin{equation}\label{RtoRsigma}
{\cal R}_v({\bf r})=\sum_{\sigma}{\cal R}_{\sigma},\quad
{\cal R}_v=\sum_{\sigma} \sigma {\cal R}_{\sigma}.
\end{equation}
Similar to (\ref{4.2}) we introduce the chiral mean densities
\begin{equation}\label{rhosigma}
\langle{\cal R}_{\sigma}({\bf r})\rangle=\rho_{\sigma}=\rho/2.
\end{equation}
The last equality follows from the mean spacial isotropy of the Berry function 
(\ref{berry}) which implies that in mean there is no preferred chirality of nodal points. 

Moreover we introduce the chiral the auxilliary correlation function
$$
  G_v(s)=\langle{\cal R}_v({\bf r}){\cal R}_v({\bf r- s})\rangle
$$  
\begin{equation}\label{Gv}
    =\left<\sum_{i,j}
 \sigma_i \delta({\bf r- r}_i)\sigma_j\delta({\bf r- r}_j-{\bf s})\right>.
\end{equation}
and correlation functions of the chiral mean densities (\ref{Rsigma}) 
$$
  G_{\sigma,\sigma'}(s)=\langle{\cal R}_{\sigma}({\bf r}){\cal R}_{\sigma'}
  ({\bf r- s})\rangle
$$
\begin{equation}\label{Gsigma}
     =\left<\sum_{i_{\sigma},j_{\sigma}'}
  \delta({\bf r- r}_{i_{\sigma}})\delta({\bf r- r}_{j_{\sigma'}}-{\bf s})\right>.
\end{equation}
From relations (\ref{3.0}), (\ref{3.1}), (\ref{4.3}), (\ref{Gv}) and (\ref{Gsigma})
it follows that
\begin{equation}\label{GtoGsigma}
  G_{\sigma \sigma'}(s)=\frac{1}{4}[G(s)+\sigma \sigma'G_v(s)].
\end{equation}

\begin{figure}
\includegraphics{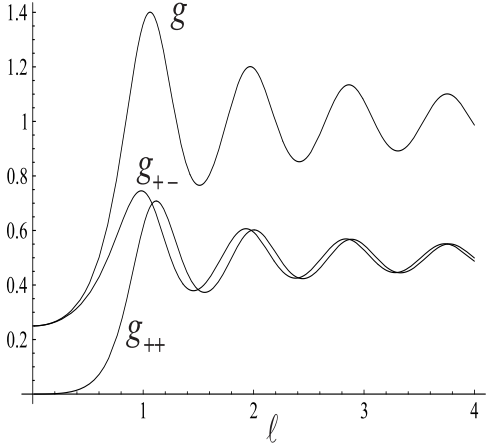}
\caption{Plots of the dimensionless correlation functions
$g_{\sigma,\sigma}(\ell)=g_+$,  $g_{\sigma,-\sigma}(\ell)=g_-$, and $g(\ell)$ 
as defined in (\ref{gsigma}).}
\label{fig5}
\end{figure}

Consider a nodal point with chirality $\sigma$. Similar to 
(\ref{4.5}) we define the mean chiral densities around this point
(\ref{rhosigma})
\begin{equation}\label{gammasigma}
\gamma_{\sigma \sigma'}(s)=\frac{2}{\rho}\;G_{\sigma \sigma'}(s).
\end{equation}
Then from (\ref{GtoGsigma}) and (\ref{gammasigma}) we obtain
\begin{equation}\label{gammachir}
  \gamma_{\sigma \sigma'}(s)=\frac{1}{2}[\gamma(s)+\sigma \sigma'\gamma_v(s)].
\end{equation}
The subscripts $(\sigma, \sigma')$ in (\ref{gammachir}) indicate that
the given point and its neighbours have the same vorticity,
while $(\sigma, -\sigma)$ refers to different chiralities for the given point
and its neighbours. 
Therefore a knowledge of the correlation functions $G(s)$ and $G_v(s)$ which are calculated 
in Appendix (formulas (\ref{56}) and (\ref{75})) is enough to find the chiral
mean densities and correlation functions.
 These chirality-dependent correlation functions are 
 shown graphically in Fig. \ref{fig5}
in the scaled form
\begin{equation}\label{gsigma}
g_{\sigma \sigma'}(\ell)=
\frac{\gamma_{\sigma \sigma'}({\ell}/\sqrt{\rho})}{\rho}.
\end{equation}
\vspace*{7cm} 
Also in this case we find good agreement with Berry and Dennis \cite{Berry3}.
Above we alluded to the general shape of  pair-correlation functions
 for amorphous 
materials. With the dependence on $\sigma$ included we may obviously carry this 
naive picture a bit further to talk in a loose way about the nodal points
as a two-component 
system in which 'objects' 
with the same vorticities (topological charges) repel each other. 
At the same time 'objects' with the same $\sigma$ may approach 
each other closely. Hence we may think about the system of 
nodal points as a thin slab of a fictitious binary amorphous solid or salt. Of course, 
this analogy should not be pushed too far.

In order to find the distributions functions
$f_{\sigma \sigma'}(\ell)$ 
for nearest distances between nodal points with chiralities $\sigma$ and 
$\sigma'$
we insert $g_{\sigma \sigma'}(\ell)$ into (\ref{dfnd2}). For the Bernoulli
approximation   one also has to 
label the mean number of points (\ref{38}) as
\begin{equation}\label{nsigma}
  \langle n_{\sigma,\sigma'}(\ell)\rangle=2\pi\int_0^\ell g_{\sigma,\sigma'}(z)z\,dz\,.
\end{equation}
which is to be be inserted in (\ref{41}).
We have found above, however, that already the Poisson approximation catches
 the gross features
of the nearest neighbor distribution. For this reason and because the Bernoulli
approximation is overly tedious we will restrict 
ourselves to the Poisson approximation at this stage. Thus we consider

\begin{equation}\label{fsigma}
  f_{\sigma \sigma'}(\ell)=2\pi\ell\,g_{\sigma \sigma'}(\ell)\,
  \exp\left(-2\pi\int_0^\ell z\,g_{\sigma \sigma'}(z)\,dz\right)\,.
\end{equation}
Using (\ref{gammachir}), (\ref{nsigma}) and (\ref{fsigma}) one can show that
the asymptote of
$ f_{\sigma,-\sigma}(\ell)$ coincides with the asymptote for the DFNDNP
(\ref{asymdfnd}). For $f_{\sigma,\sigma}(\ell)$, however,
we obtain the quite different form
\begin{equation}\label{50}
 f_{\sigma,\sigma}(\ell)=\frac{5\pi}{81}\sqrt{10\pi}\ell^4 \approx
 1.09 \ell^4.
\end{equation}

\begin{figure}
\includegraphics{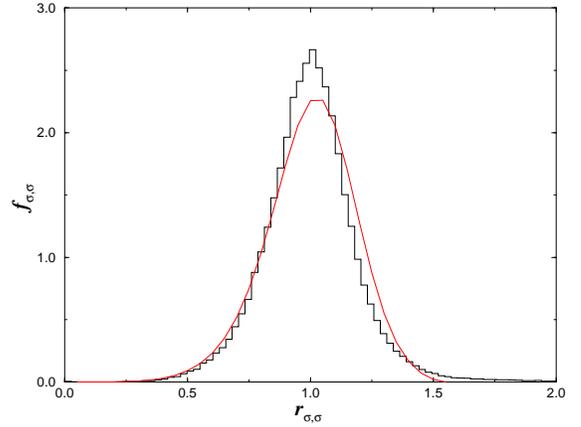}
\caption{Distribution functions of nearest
distances between nodal points with the same chirality $\sigma$ versus
dimensionless distance (\ref{32}) with $<\ell>=0.998$.
Solid line is calculated from function (\ref{fsigma}).  The histogram refers to numerical
work as described in Fig. \ref{fig2}. For this case we obtain the same value
 for $<\ell>$.}
\label{fig6}
\end{figure}

\begin{figure}
\includegraphics{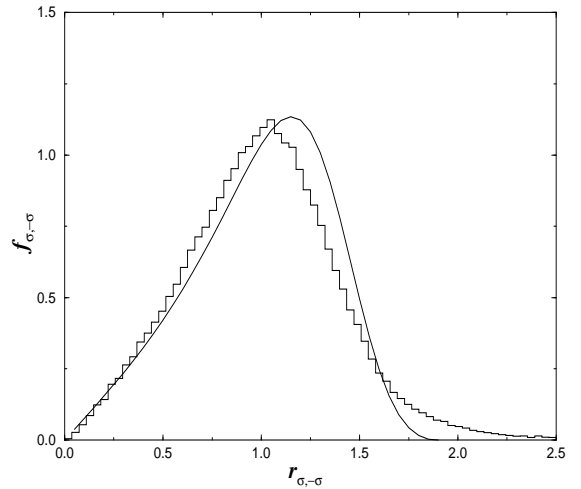}
\caption{The same as in Fig.\ref{fig6} but for nodal points
with opposite chiralities $\sigma$; $<\ell> = 0.696$.
The histogram refers to numerical
work as described in Fig. \ref{fig2}. For this case we find $<\ell> = 0.726$.}
\label{fig7}
\end{figure}

The $\sigma$-dependent DFNDNP (\ref{fsigma}) for $(\sigma,\sigma)$ and
$(\sigma,-\sigma)$ are compared with numerical 
results in Figs \ref{fig6} and \ref{fig7}. The difference between
the two combinations $(\sigma,\sigma)$ and $(\sigma,-\sigma)$ is very clear.
In the first case there is a strong repulsion between the nearest neighbors, 
a result that is to be expected from Fig. \ref{fig5}. The distribution is, however, of a very 
simple form. Essentially is corresponds a symmetric ring ('first shell') 
around the given point. Except for the low tail regions it is well 
approximated by a Gaussian, i.e., there is basically a random distribution 
within the 'shell'. 
\vspace*{65mm} 

\vspace*{65mm} 
In the second case, ($\sigma,-\sigma$), the distribution shows how the 
nodal points are allowed to come arbitrary close to each other and how the shell
 structure is less pronounced. For small separations $f_{\sigma,-\sigma}$ 
 is obviously the dominant term in the total distribution $f$. 
 Finally we note that $<\ell> \simeq 0.7$ for opposite chiralities 
 (topological charges) and $<\ell> \simeq 1$ for equal chiralities, i.e., there
 are  'inner' and  'outer shells' for  
 opposite and equal vorticities, 
 respectively. As shown by Figs \ref{fig6} and \ref{fig7} the Poisson 
 approximation reproduces the numerical results in 
 at least a qualitatively correct way.

\section{Summary}
We have considered the distribution among nodal points associated with chaotic
wave dynamics. The nodal points are of special interest as they are 
associated with vortical flow and chirality $\sigma$ which is either $+1$ or $-1$.
 In particular we have focused on the distribution of nearest separations among
 the nodal points. The reason is that distributions of this kind may carry 
 information about chaotic dynamics as conjectured in \cite{Berg1}.
 We have introduced analytic approaches based on 
complex Gaussian random functions with the known correlation
function $J_0(ks)$. Two cases have been considered, namely the Poisson
 and Bernoulli approximations. 
 
 As a supplement to the analytic approaches 
 we have performed numerical calculations to locate the nodal points and their
 vorticity using the Berry chaotic wave function in eq. (\ref{berry}). 
 The numerical distributions computed in this way are in principle 
 the correct ones and are therefore
 useful for testing the accuracy of the analytic modeling. 
We thus find that already the Poisson approximation gives a good qualitative 
understanding of the distribution of nearest neighbor separations. Some higher
 order correlations are, however, not accounted for. For this reason we have also
 considered the Bernoulli approximation which picks up some of these
 features. On the other hand the behaviour at large distances needs further work.
 
 When the distribution of nearest neighbor distances is separated into  
 distances between nodal points with equal and opposite chiralities  
 one obtains a picture that superficially reminds of binary amorphous matter.
 There are distinct 'inner' and 'outer shells'
  associated with opposite and equal chiralities, respectively. While points
  with opposite chiralities may get close to each other there is a strong repulsion
  among pairs with equal chirality. The reason is, loosely speaking, that nodal points with equal
  chirality have to appear in conjunction with 'anti-vortices' or saddle points. 
  These 'anti-vortices' act as local 'beam splitters' and are necessary
  for two
  nearby vortices to spin the same way. These 
  interesting features should be pursued further because the 
  distributions among the different phase singularities are obviously 
  interrelated. Effectively we may then arrive at a description reminding of a 
  ternary amorpous materials.
  
  The distributions discussed here relate 
  to generic features of a wave-chaotic state. For this reason it would be of interest
  if they could be verified experimentally. Using, e.g., micro-wave cavities 
  this appears to be a real possibility \cite{stoeck}.

\section*{Acknowledgements}
This work was supported by the Swedish Institute (A.I.S.), 
the Royal Swedish Academy of Sciences (A.F.S.), 
 the Krasnoyarsk 
Science Foundation (A.F.S., Grant 9F82), and the Swedish Natural 
Science Research Council. We are also grateful to
 Michael Berry and Mark Dennis for discussions and for making refs. 
 \cite{Berry3}
 and \cite{Berry4} available to us before publication. The problem of
 the statistical properties of vortices in 2D is somewhat esoteric and
  has remained dormant for many years. We are all amazed that 
  we have chosen to work on this problem more or less in parallel.

\appendix
\section*{Correlation functions for the nodal points}
In this appendix we outline the somewhat tedious derivation of the
various correlation functions.
Consider the correlation function of the random density (\ref{4.3})
\begin{eqnarray}\label{51}
G(s)=\qquad \qquad \qquad \qquad \qquad \nonumber\\
\langle|\omega({\bf r})\omega({\bf r+s})|
\delta(u({\bf r}))\delta(u({\bf r+s})
\delta(v({\bf r}))\delta(v({\bf r+s})
\rangle\,.
\end{eqnarray}
Since $u$ and $v$ are statistically independent random fields it is sufficient
to consider only the statistical properties of field $u$.
First of all, note that the joint distribution of the values of the scalar
Gaussian random field at the points ${\bf r}$ and ${\bf r+s}$ has the
form
$$
W(u,u_s)=\langle\delta(u({\bf r})-u)\delta(u({\bf r+s})-u_s)
\rangle
$$
\begin{equation}\label{52}
=\frac{1}{2\pi\sqrt{1-a^2(s)}}
\exp\left[-\frac{u^2+u_s^2-2a(s)u u_s}{2[1-a^2(s)]}\right]\,.
\end{equation}
Futhermore, we will need the reciprocal
statistical properties of Gaussian scalar field $u$ and its
gradient $\nabla u$. They are completely defined by the correlation
vector
\begin{equation}\label{53}
{\bf e(s)}=\langle u({\bf r+s})\nabla u({\bf r}) \rangle
\end{equation}
and the correlation tensor $b_{ij}({\bf s}),\,i,j=1,2$ of vector
field $\nabla u$.

It is convenient to project this vector and
tensor onto the local coordinate system  related to the
vector ${\bf s}$ through the longitudinal components
$$
u_\parallel({\bf r}), \quad u_\parallel({\bf r+s})
$$
and the transverse ones
$$
u_\perp({\bf r}), \quad u_\perp({\bf r+s}).
$$
These components have the remarkable correlation properties
\begin{equation}\label{54}
\langle u({\bf r+s}) u_\parallel({\bf r})\rangle=e(s)\,,\quad
\langle u({\bf r+s}) u_\perp({\bf r})\rangle =0\,.
\end{equation}
The tensor correlation function becomes diagonal
\begin{equation}\label{55}
\langle u_\parallel({\bf r}) u_\parallel({\bf r+s})\rangle=
b_\parallel(s)\,\quad
\langle u_\perp({\bf r}) u_\perp({\bf r+s})\rangle=
b_\perp(s)\,,
\end{equation}
$$
\langle u_\parallel({\bf r})u_\perp({\bf r+s})\rangle= 0\,.
$$
Here we introduce the following  notations
$$
b_\parallel(s)=-\frac{d^2 a(s)}{ds^2}\,,\quad
b_\perp(s)=-\frac{1}{s}\frac{d a(s)}{ds}\,,\quad
e(s)=-\frac{d a(s)}{ds}\,,
$$
$$
b_\parallel (0)=b_\perp (0)=\frac{1}{2}k^2=b\,.
$$
We may then write the density correlation function (\ref{51})
as
\begin{equation}\label{56}
G(s)=\frac{1}{4\pi^2[1-a^2(s)]}
\langle|u_\parallel v_\perp-v_\parallel u_\perp|
|u_{\parallel s}v_{\perp s}-v_{\parallel s}u_{\perp s}|\rangle .
\end{equation}
For brevity  we have here written the dependence on $s$ as an index.
The following averages are performed for the eight manifold
Gaussian fields
\begin{eqnarray}\label{57}
\{u_\parallel({\bf r}), u_\parallel({\bf r+s})\}, \quad
\{v_\parallel({\bf r}), v_\parallel({\bf r+s})\}\nonumber\\
\{u_\perp({\bf r}), u_\perp({\bf r+s})\}, \quad
\{v_\perp({\bf r}), v_\perp({\bf r+s})\}
\end{eqnarray}
Here each pair of variables has the correlation properties
\begin{eqnarray}\label{58}
\langle u_\parallel^2\rangle=\langle v_\parallel^2\rangle
=b\Delta(s), \nonumber\\
\langle u_\perp^2\rangle=\langle v_\perp^2\rangle=b,\nonumber\\
\langle u_\parallel u_{\parallel s} \rangle=
\langle v_\parallel v_{\parallel s} \rangle=c(s),\nonumber\\
\langle u_\perp u_{\perp s} \rangle=\langle v_\perp v_{\perp s} \rangle
=b_\perp(s)
\end{eqnarray}
where
\begin{eqnarray}\label{59}
\Delta(s)=\frac{b[1-a^2(s)]-e^2(s)}{b[1-a^2(s)]}\,,\nonumber\\
c(s)=\frac{b_\parallel(s)[1-a^2(s)]-a(s)e^2(s)}{b[1-a^2(s)]}\,.
\end{eqnarray}
It is convenient to transform to normalized random variables
$$
\tilde{u}_\parallel=\frac{u_\parallel}{\sqrt{b\Delta(s)}},\quad
\tilde{v}_\parallel=\frac{v_\parallel}{\sqrt{b\Delta(s)}},\quad
\tilde{u}_\perp=\frac{u_\perp}{\sqrt{b}},\quad
\tilde{v}_\perp=\frac{v_\perp}{\sqrt{b}}
$$
With the correlation coefficients
\begin{equation}\label{60}
\alpha=\frac{b_\parallel(s)[1-a^2(s)]-a(s)e^2(s)}
{b[1-a^2(s)]-e^2(s)},\quad
\beta=\frac{b_\perp(s)}{b}.
\end{equation}
 expression (\ref{56}) now takes the form
\begin{equation}\label{61}
G(s)=\rho^2\frac{\Delta(s)}{1-a^2(s)}\Lambda(\alpha,\beta)
\end{equation}
where
\begin{equation}\label{62}
\Lambda(\alpha,\beta)=
\langle|\tilde{u}_\parallel \tilde{v}_\perp-
\tilde{v}_\parallel \tilde{u}_\perp|
|\tilde{u}_{\parallel s}\tilde{v}_{\perp s}-
\tilde{v}_{\parallel s}\tilde{u}_{\perp s}|\rangle
\end{equation}
and the averaging is performed with respect to the distributions
\begin{eqnarray*}
w_\parallel(u,u_{s})=
\frac{1}{2\pi\sqrt{1-\alpha^2}}\exp\left[-\frac{u^2+u_{s}^2-
2\alpha u u_{s}}{2(1-\alpha^2)}\right]\nonumber\\
w_\perp(u,u_{s})=
\frac{1}{2\pi\sqrt{1-\beta^2}}\exp\left[-\frac{u^2+u_{s}^2-
2\beta u u_{s}}{2(1-\beta^2)}\right].
\end{eqnarray*}
Here distribution $w_\parallel$ describes the statistics of
``parallel pairs'' $\{u_\parallel,u_{\parallel s}\}$ and
$\{v_\parallel,v_{\parallel s}\}$ while $w_\perp$ describes
properties of ``perpendicular'' ones.

It remains to calculate the function (\ref{62})  which will be done
in two steps. Firstly
we average over the statistical ensemble of parallel components
$\{u_\parallel,u_{\parallel s},v_\parallel,u_{\parallel s}\}$
at all given perpendicular variables to obtain
\begin{equation}\label{63}
\Lambda_\perp(\alpha,\beta)=\langle|\lambda\lambda_s|\rangle_\perp
\end{equation}
where notation $\langle\cdots\rangle_\perp$ means that all
perpendicular variables are considered as fixed, and that
$$
\lambda=
\tilde{u}_\parallel \tilde{v}_\perp-
\tilde{v}_\parallel \tilde{u}_\perp,\quad
\lambda_s=\tilde{u}_{\parallel s}\tilde{v}_{\perp s}-
\tilde{v}_{\parallel s}\tilde{u}_{\perp s}
$$
are two Gaussian variables with zero mean values with dispersions
$$
\langle\lambda_\perp^2\rangle=
\tilde{u}_\perp^2+\tilde{v}_\perp^2,\quad
\langle\lambda_{\perp s}^2\rangle=
\tilde{u}_{\perp s}^2+
\tilde{v}_{\perp s}^2
$$
and correlation
$$
\langle\lambda\lambda_s\rangle_\perp=
\alpha (\tilde{u}_\perp \tilde{u}_{\perp s}+
\tilde{v}_\perp \tilde{v}_{\perp s}).
$$
Let us  normalize variables $\lambda$ and $\lambda_s$ by the relations
$$
z=\frac{\lambda}{\sqrt{u_\perp^2+v_\perp^2}},\quad
z_s=\frac{\lambda_s}{\sqrt{u_{\perp s}^2+v_{\perp s}^2}}
$$
which transform the function (\ref{63})
as
\begin{equation}\label{64}
\Lambda_\perp(\alpha,\beta)=\sqrt{(u_\perp^2+v_\perp^2)
(u_{\perp s}^2+v_{\perp s}^2)}\,\langle|z z_s|\rangle_\perp
\end{equation}
where $z$ and $z_s$ are Gaussian
variables with unit dispersion and the correlation
coefficient
as
$$
\gamma=\langle z z_s\rangle_\perp=\alpha\frac{u_\perp u_{\perp s}+
v_\perp v_{\perp s}}
{\sqrt{(u_\perp^2+v_\perp^2)(u_{\perp s}^2+v_{\perp s}^2)}}.
$$

Using the properties of Gaussian random variables one can derive after
some algebra the following equation
\begin{eqnarray}\label{65}
\frac{d^2\langle|z_1 z_2|\rangle_\perp}{d\gamma^2}=
4\langle\delta(z_1)\delta(z_2)
\rangle_\perp\nonumber\\
=\frac{2}{\pi \sqrt{1-\gamma^2}}.
\end{eqnarray}
Taking into account the initial conditions
$$
\langle|z_1 z_2|\rangle_\perp\bigr|_{\gamma=0}=\frac{2}{\pi},\quad
\frac{d\langle|z_1 z_2|\rangle_\perp}{d\gamma}\bigr|_{\gamma=0}=0
$$
we obtain from (\ref{65})
\begin{equation}\label{66}
{\cal F}(\gamma)=\langle|z z_s|\rangle_\perp=\frac{2}{\pi}
\left[\sqrt{1-\gamma^2}+\gamma\arcsin\gamma\right].
\end{equation}
Substituting (\ref{66}) into (\ref{64}) and averaging
over the remaining four perpendicular random variables
we obtain
\begin{equation}\label{67}
\Lambda(\alpha,\beta)=
\langle\Lambda_\perp(\alpha,\beta)\rangle=\langle\sqrt{
(u_\perp^2+v_\perp^2)(u_{\perp s}^2+v_{\perp s}^2)}{\cal F}(\gamma)\rangle.
\end{equation}
The angular brackets on the right hand of this equation imply an
averaging over the $\perp$-variables with
the following joint distribution
\begin{eqnarray*}
w(u,u_s,v,v_s)=\frac{1}{4\pi^2(1-\beta^2)}\nonumber\\
\times\exp\left[-\frac{u^2+v^2+u^2_s+v_s^2-2\beta(u u_s+v v_s)}
{2(1-\beta^2)}\right].
\end{eqnarray*}
In order to perform integration over four $\perp$ variables
$\{u,u_s,v,v_s\}$  we use the polar system of coordinates
$$
u_\perp=\xi\cos\varphi,\quad v_\perp=\xi\sin\varphi\quad
u_{\perp s}=\eta\cos\psi,\quad v_{\perp s}=\eta\sin\psi
$$
which gives, instead of (\ref{67}),
\begin{equation}\label{68}
\Lambda(\alpha,\beta)=
\langle\xi\eta\mathcal{F}(\alpha\cos\mu)\rangle
\quad (\mu=\varphi-\psi).
\end{equation}
Here the three random variables $\{\xi,\eta,\mu\}$ are distributed  as
$$
w(\xi,\eta,\mu)=\frac{\xi\eta}{2\pi(1-\beta^2)}\exp\left[-
\frac{\xi^2+\eta^2-2\beta\xi\eta\cos\mu}{2(1-\beta^2)}\right],
$$
$$
\quad\xi,\,\eta>0,\quad \mu\in [-\pi,\pi].
$$
\noindent To exclude the random variables $\xi$ and $\eta$ we
rewrite relation (\ref{68}) as
\begin{equation}\label{69}
\Lambda(\alpha,\beta)=\frac{1}{2\pi}\int_{-\pi}^\pi
{\cal F}(\alpha\cos\mu){\cal A}(\mu,\beta)d\mu
\end{equation}
where
\begin{eqnarray}\label{70}
{\cal A}(\mu,\beta)=\frac{1}{1-\beta^2}\qquad \qquad \qquad\nonumber\\
\times\int_0^\infty d\xi\int_0^\infty d\eta\;
\xi^2\eta^2 \exp\left[-\frac{\xi^2+\eta^2-
2\beta\xi\eta\cos\mu}{2(1-\beta^2)}\right]\,.
\end{eqnarray}

If we use the new variables of integration $p,\delta$ defined through
$$
\xi=p\cos\delta\,,\quad \eta=p\sin\delta
$$
the integral (\ref{70}) transforms into the form
$$
{\cal A}(\mu,\beta)=\frac{1}{8}(1-\beta^2)^2
$$
$$
\int_0^\pi d\theta\sin^2\theta\int_0^\infty dp\; p^5\,
\exp\left[-\frac{1}{2}p^2(1-\beta\cos\mu\sin\theta)\right]
$$
with  $\theta=2\delta$.
After integration over $p$ we obtain
\begin{equation}\label{71}
{\cal A}(\mu,\beta)=
(1-\beta^2)^2 Q (\beta\cos\mu).
\end{equation}
\begin{equation}\label{72}
Q(z)=
\frac{3z}{(1-z^2)^2}+\frac{1+2z^2}{(1-z^2)^{5/2}}
\left[\arctan\left(\frac{z}{\sqrt{1-z^2}}\right)+\frac{\pi}{2}
\right].
\end{equation}
Substituting (\ref{71}) into (\ref{69}) we finally obtain
\begin{equation}\label{73}
\Lambda(\alpha,\beta)=\frac{(1-\beta^2)^2}{2\pi}\int_{-\pi}^\pi
{\cal F}(\alpha\cos\mu) Q(\beta\cos\mu)d\mu
\end{equation}
where the function ${\cal F}(\gamma)$ is defined by the equality
(\ref{66}) and the function $Q(\gamma)$  is given by (\ref{72}).

For the $\sigma$-dependent density correlation functions
it is easy to show, using the Gaussian properties of the fields
$u$ and $v$, that the correlation function for $\sigma$-dependent random
density (\ref{3.1}) has the form
\begin{equation}\label{75}
G_v(s)=\langle R_v({\bf r}) R_v({\bf r+ s})\rangle
  =\rho^2 \frac{2\Delta(s)}{1-a^2(s)}\alpha(s)\beta(s).
\end{equation}

\end{document}